\documentclass[11pt,reqno]{amsart} 
\usepackage{comment}
\usepackage{amsmath}
\usepackage{amssymb}
\usepackage{url}
\usepackage{amsfonts}
\usepackage[parfill]{parskip}  
\usepackage[foot]{amsaddr}

\usepackage{mathabx}

\newcommand{\Inf}{\operatorname{Inf}}
\newcommand{\InfSq}{\operatorname{Inf^{\,\rm{sq}  } }}
\newcommand{\infsq}{\operatorname{Inf^{\,\rm{sq}  } }}

\usepackage{amsmath,amssymb,amsthm}    
\usepackage{graphicx}

\usepackage{hyperref} 

\makeatletter
\def\thm@space@setup{%
  \thm@preskip=\parskip \thm@postskip=0pt
}
\makeatother  

\usepackage{amsfonts}
\usepackage{fullpage}
\newtheorem{theorem}{Theorem}[section]

\newtheorem{lemma}[theorem]{Lemma}
\newtheorem{proposition}[theorem]{Proposition}

\newtheorem{fact}[theorem]{Fact}

\theoremstyle{definition}
\newtheorem{definition}[theorem]{Definition}

\newtheorem{notation}[theorem]{Notation}
\newtheorem{remark}[theorem]{Remark}

\newcommand{\R}{{\mathbb R}}

\DeclareMathOperator*{\EX}{\mathbb  E}

\newcommand{\Ex}{ \EX}
\DeclareMathOperator{\sgn}{sgn}

\usepackage{xy}
\xyoption{matrix}
\xyoption{frame}
\xyoption{arrow}
\xyoption{arc}

\usepackage{ifpdf}
\ifpdf
\else
\PackageWarningNoLine{Qcircuit}{Qcircuit is loading in Postscript mode.  The Xy-pic options ps and dvips will be loaded.  If you wish to use other Postscript drivers for Xy-pic, you must modify the code in Qcircuit.tex}
\xyoption{ps}
\xyoption{dvips}
\fi

\entrymodifiers={!C\entrybox}

\usepackage{setspace}

\usepackage{hyperref}

\usepackage{fullpage}

\newcommand{\MS}{\mathcal M_S^{ \scriptscriptstyle =1} }
\newcommand{\Ms}{\mathcal M^{ \scriptscriptstyle =1} }

\providecommand{\intd}{\mathrm{d}}

\DeclareMathOperator{\Res}{Res}
\begin{document}

\title{On the sum of $L_1$ influences}

\author[A. Ba\v ckurs]{Art\=urs Ba\v ckurs}
\email{backurs@mit.edu }
\author[M. Bavarian]{Mohammad Bavarian } 
\email{bavarian@mit.edu }
\address{Massachusetts Institute of Technology, Cambridge, MA, USA. }

\date{}

\maketitle

\begin{abstract}

For a function $f$ over the discrete cube, the total $L_1$ influence of $f$ is defined as $\sum_{i=1}^n \|\partial_i  f\|_1$, where $\partial_i f$ denotes the discrete derivative of $f$ in the direction $i$. In this work, we show that the total $L_1$ influence of a $[-1,1]$-valued function $f$ can be upper bounded by a polynomial in the degree of $f$, resolving affirmatively an open problem of Aaronson and Ambainis (ITCS 2011). 
The main challenge here is that the $L_1$ influences do not admit an easy Fourier analytic representation.  In our proof, we overcome this problem by introducing a new analytic quantity $\mathcal I_p(f)$, relating this new quantity to the total $L_1$ influence of $f$. This new quantity, which roughly corresponds to an average of the total $L_1$ influences of some ensemble of functions related to $f$, has the benefit of being much easier to analyze, allowing us to resolve the problem of Aaronson and Ambainis. We also give an application of the theorem to graph theory, and  discuss the connection between the study of bounded functions over the cube and the quantum query complexity of partial functions where Aaronson and Ambainis encountered this question.

\end{abstract}

\maketitle
\section{Introduction}

The notion of the influence of a variable \cite{BKKL, KKL} plays a fundamental role in the study of functions over product probability spaces. A canonical example of a product probability space is the discrete cube $\{-1,1\}^n$ equipped with the uniform probability measure. Given a Boolean function $f:\{-1,1\}^n\rightarrow \{-1,1\}$, the i\textsuperscript{th} influence of $f$ is defined as  the fraction of the edges in the direction i  where  the value of $f$ changes along the edge, i.e. 
\[ \Inf_i(f):= \Pr_{x\in\{-1,1\}^n} \left[ f(x)\neq f(x^i) \right]. \]
Here $x^i$ denotes the neighbor of the point $x$ in the i\textsuperscript{th} direction that is   \[x^i=(x_1,\ldots, x_{i-1}, -x_i,x_{i+1},\ldots, x_n).\]
The sum over all the influences  is called the total influence, $\Inf(f)=\sum_{i=1}^n \Inf_i(f)$, and has a clear geometric meaning, as the total edge boundary between the set of points where $f=1$ and its complement.

Going beyond the Boolean valued functions, the notion of the influence of a variable can be generalized in several ways. The idea is to replace the term $\Pr[ f(x)\neq f(x^i)]$ with an analytical expression such as $\Ex\left[ \left(\frac{|f(x)-f(x^i)|}{2}\right) ^\alpha\right]$ for some non-zero $\alpha\in \R$.\footnote{Notice that for a $\{-1,1\}$-valued function $f$, the term $|f(x)-f(x^i)|/2$ is either $0$ or $1$; hence all these different notions of influence of a variable coincide in this setting.
.} Two important cases to consider are: $\alpha=1$ and $\alpha=2$, which correspond to the $L_1$ and $L_2$ influences respectively.  Since we are concerned mostly with the $L_1$ case here, following Aaronson and Ambainis \cite{a}, we take up the following notation (for the notation see Section \ref{prelim}.).
\begin{definition}
Given a function $f:\{-1,1\}^n\rightarrow \R$, we denote the i\textsuperscript{th} discrete derivative of $f$ by $\partial_i f(x)= (f(x)- f(x^i))/2$. We denote the i\textsuperscript{th} $L_1$ influence of $f$ by $\Inf_i(f)= \| \partial_i f \|_1$ and the total $L_1$ influence by  $\Inf(f)=\sum_{i=1}^n \| \partial_i f\|_1$. Note that the Fourier expansion of $\partial_i f$ can be recovered from that of $f$ as follows:
\[ \partial_i f= \sum_{S\ni i} \hat{f}(S) \chi_S .\]

\end{definition}

In \cite{a}, Aaronson and Ambainis asked whether the total $L_1$ influence of a $[-1,1]$-valued function can be bounded  in terms of a polynomial of the function's Fourier degree. In this work, we resolve their question affirmatively. 
\begin{theorem}
\label{th:1}
Let $f:\{-1,1\}^n\to \R$ and $\deg(f)=d$. Then we have 
\[
\Inf(f)=O( d^3  \|f\|_\infty ).
\]
\end{theorem}
Aaronson and Ambainis did not conjecture anything about the growth of that polynomial. It is likely that in fact a linear bound $O(d)$ is sufficient for the above result to hold.
\begin{remark} Theorem \ref{th:1} is most interesting when $d$ is small compared to $n$, as it is straightforward to see that $\Inf(f)= O(n \|f\|_\infty)$. Hence, for concreteness it might be useful to think of $d$ as $n^{o(1)}$ or even a large constant. 
\end{remark}
We should mention that we are not aware of any bound independent of $n$ (say $\exp(d)$) which would easily follow from the previous results in the literature. 

To build intuition about the theorem, it is useful to first consider the analogous question over  $L_2$. The total $L_2$ influence of a function $f:\{-1,1\}^n\rightarrow \R$ is  defined as the sum over all directional $L_2$ influences, i.e., \[\infsq(f)=\sum_{i=1}^n \Inf_i^{\rm{sq}}(f)=\frac{1}{4}\sum_{i=1}^n \Ex[ (f(x)-f(x^i))^2].\] One reason for considering the $L_2$ case  is that the self-duality of $L_2$ allows one to derive a nice characterization of the $L_2$ influences as the average weight of Fourier coefficients. More precisely, we have
\begin{equation}
 \InfSq(f)= \sum_{S\subseteq[n]} |S| \, \hat{f}(S)^2. \qquad \label{l2:influence}  
\end{equation}
The dual expression (\ref{l2:influence}) for the total $L_2$ influence  leads to a quick proof of an $L_2$ analogue of Theorem \ref{th:1}, which is
\begin{equation}
 \InfSq(f) \leq \deg(f) \|f\|_2^2 . \label{l2:ineq}  
 \end{equation}
Indeed, Aaronson and Ambainis's question was based partly on the empirical evidence, and partly on the fact that the similar statement holds in the $L_2$ setting. However, it turns out that proving the analogous statement in the $L_1$ case is much more difficult. The main difficulty is due to the fact that unlike the $L_2$ influences, the $L_1$ influences do not have an easy characterization in terms of Fourier coefficients.  Hence, to prove Theorem \ref{th:1}, one needs to relate the $L_1$ influences of a function, which is defined in terms of the values of the function, to its degree that is most easily understood in terms of the values of its Fourier coefficients. Notice that this difficulty was not present in  the $L_2$ case because of the dual characterization of $L_2$ influences in equation (\ref{l2:influence}), which has no analogue in any setting beyond $L_2$.

\subsection*{Techniques and the organization of the paper}
Beside the main contribution of this paper, which is the proof of Theorem \ref{th:1}, we believe the second significant contribution of this work lies in the new definitions and techniques introduced on the way to  the final result. The main technical machinery used in the proof of Theorem \ref{th:1} is presented in Section \ref{sec:proof-overview}. There we define and analyze an  auxiliary operator $\MS$, whose properties allow us to relate the total $L_1$ influence to a more tractable quantity $\mathcal I_p(f)$.
The above operators and quantities turn out to be quite natural from a mathematical point of view, and understanding 
their properties could be useful elsewhere. Once the auxiliary quantities and operators are introduced and their properties investigated, Theorem \ref{th:1} follows rather quickly. 
Almost all the main ingredients of the proof are already present in Section \ref{sec:proof-overview}; the missing technical details are presented in Section \ref{sec:homg} and \ref{sec:general} and in the appendix to complete the proof. 

After establishing Theorem \ref{th:1}, we apply this theorem to give a new proof of the following theorem of Erd\H{o}s, Goldberg and Pach from graph theory:
\begin{theorem}[Erd\H{o}s et al.] \label{application_thm}
Given a graph $G=(V,E)$ with density $\rho_G= |E|/\binom{n}{2}$, there always exists a cut $(S,S^c)$ such that 
\[ \left| E(S, S^c)-\rho_G |S||S^c| \right|=\Omega\left(\min(\rho_G, 1-\rho_G)\, n^\frac{3}{2}\right). \]
\end{theorem}
The above is proved by applying Theorem \ref{th:1} to 
\begin{align*} g_G(x)&:=\frac{|E|}{2}-\rho_G\frac{|V|(|V|-1)}{4} \\
 &+\frac{\rho_G}{2} \sum_{i<j} \,x_i x_j - 
\left(\frac{1}{2}-\frac{\rho_G}{2}\right) \sum_{(i,j)\in E}\,  x_i x_j. \end{align*}
The above example demonstrates that in some settings Theorem \ref{th:1} can be much stronger than its $L_2$ counterpart --- i.e. equation (\ref{l2:influence}). To see this, note that when $G$ is  a random graph of some fixed constant density (say $\rho_G=1/2$), the $\Omega(n^{3/2})$ bound in Theorem \ref{application_thm}, proved via Theorem \ref{th:1}, is tight. On the other hand, applying the $L_2$ bound of equation (\ref{l2:influence}) to $g_G$, one only gets an inferior lower bound of $\Omega(n)$.\footnote{Let us elaborate (see also Section \ref{sec:app}) the polynomial $g_G$ is chosen such that $\|g_G\|_\infty$ is equal to the maximal cut deviation of 
the graph $G$. Applying $\text{max-cutdev}(G)=\|g_G\|_\infty \geq \|g_G\|_2= \Omega( \infsq(g_g)^{\frac{1}{2}} )$ 
only gives us $\Omega(n)$ bound. Essentially, this is due to the fact that $\|g_G\|_2$ here is much smaller than $\|g_G\|_\infty$. Indeed, in this setting, equation (\ref{l2:influence}) seems more suitable for capturing the average case cut-deviation behavior, whereas Theorem \ref{th:1} seems better at capturing extremal cases.}


\section{Related work and background}\label{sec:related_work} 
The importance of the concept of influences in the analysis of functions over product spaces was already recognized in the pioneering work of Kahn, Kalai and Linial \cite{KKL} and Bourgain et al. \cite{BKKL}. Building upon these results, Friedgut \cite{Friedgut} showed that a Boolean function with very low total influence is somewhat ``simple" as it can be approximated with a function depending on few coordinates, showing that the total influence in some regime acts as a complexity measure of functions. Bourgain \cite{Bourgain} further studied the interaction between the condition of Boolean-valuedness and influences proving very powerful results about the \emph{spectrum} of such functions.\footnote{The spectrum of a function usually refers to the weight distribution of Fourier coefficients of a function. The spectral properties refer, for example, to the behavior of $S(m)=\sum_{|S|\geq m} |\hat{f}(S)|^2$ or $R(k)=\sum_{S\subseteq [n]} |S|^k |\hat{f}(S)|^2$ as a function of $m$ and $k$.} Later, Dinur et al. \cite{Dinur} obtained a (exponentially weaker but optimal) generalization of Bourgain's result \cite{Bourgain} for the $[-1,1]$-valued functions.

Most of the results mentioned above, either implicitly or explicitly, investigated the effects of Boolean-valuedness on the spectrum of functions. As most computational and learning problems are specified by a truth table of the form $f:\{-1,1\}^n\rightarrow \{0,1\}$, one may assume that understanding the spectral properties of Boolean functions should be sufficient for the applications to complexity theory. Indeed, for many applications such as the study of small-depth circuits, threshold circuits, decision trees and even, via an easy reduction, the bounded error query complexity of total functions, this is sufficient. The main point here is the distinction between \emph{total functions} versus \emph{partial functions}. A total function is a Boolean function $f:\{-1,1\}^n\rightarrow \{0,1\}$ defined on the whole hypercube whereas a partial function $f:A\rightarrow \{0,1\}$ is only defined on a strict subset $A$ of the hypercube. The distinction between partial functions and total functions is crucial in the applications to query and communication complexity. For example, although it has been known since the work of Simon \cite{simon} that for partial functions quantum algorithms can be exponentially more powerful than classical algorithms, for total functions quantum algorithms can only exhibit at most a polynomial speedup. (See \cite{be, buhrman_wolf} for further discussion and  \cite{Gavinsky, Regev, wolf} for similar issues in communication complexity.) 

It turns out that the case of quantum and randomized query complexity of partial functions is much less understood than that of total functions. The work of Aaronson and Ambainis \cite{a} is one of the first papers trying to investigate the relationship between the size and the structure of the domain $A$ of a partial function $f$, and the quantum versus classical advantage achievable for computing $f$. The intuition is that unless the domain $A$ is specially structured and rather small, quantum algorithms should not be able to outperform classical algorithms by much. Unfortunately, our knowledge in this topic is still quite limited. 

One of the first complications that arises when trying to address the problems regarding the query complexity of partial functions is that, instead of Boolean functions, one has to deal with more general bounded functions. To see this, let us first recall how one usually associates a polynomial to any (say, quantum) algorithm solving a query problem.


\begin{lemma}[See \cite{be}]
\label{poly}
Let $Q$ be a quantum algorithm with a black box access to an input $X\in \{-1,1\}^n$, trying to solve a problem $f:\{-1,1\}^n\rightarrow \{0,1,*\}$. \footnote{ This is the alternative notation for partial query complexity problems with $A={\rm{dom}}(f)$, consisting of points $x$ where $f(x)\neq *$. We say that an algorithm accepts an input if it outputs $1$ and it rejects an input if it outputs $0$.} If $Q$ makes $T$ queries to the black-box before accepting or rejecting the input, its acceptance probability of each $X\in \{-1,1\}^n$ can be seen as a real-valued multilinear polynomial $p(X)$ of degree at most $2T$.
\end{lemma}
Hence, we see that if an algorithm manages to solve a query problem in few queries, this implies the existence of a polynomial $p(X)$ of low degree satisfying $|p(X)- f(X)|\leq 1/3$ for any $X$ in the domain $A$ of $f$. Hence, if the domain of $f$ is a strict subset of Boolean hypercube, one has no information on $p(X)$ for $X\in A^c$. Unlike the case of essentially Boolean functions, i.e. functions with the range $[0,1/3]\cup [2/3,1]$ which in many respects resemble Boolean functions in their spectral behavior (in fact, many of the same techniques used for studying Boolean functions also apply here), the spectral properties of bounded functions can be quite different from those of Boolean functions as demonstrated by the work of Dinur et al. \cite{Dinur}. Thus it seems that one prerequisite for making progress on problems regarding the tradeoffs between the size and the structure of the domain of a partial function and the quantum and classical query complexity is to develop more analytical tools for studying the properties of bounded functions over the hypercube. The new results and techniques in this paper are precisely along such lines.


\subsection*{Improved bounds and subsequent work}
We shall note that an earlier version of this paper proved slightly inferior bounds of $O(d^3\log d)$ for Theorem \ref{th:1} and $O(d^2 \log d)$ for Theorem \ref{thm:homg}. The extra logarithmic factor in the bounds arose from a suboptimal construction of the measure $H$ in Lemma \ref{lem:vander}. In that same early version, We suggested that the extra logarithmic factor could perhaps be removed by a more careful choice of the measure $H$ concentrated on the roots of Chebyshev polynomials (as opposed to an arithmetic progression which was the basis of the original construction). Y. Filmus in fact succeeded in constructing such a measure based on  the roots of Chebyshev polynomials; he communicated the proof to us and he kindly allowed us to include it in this paper. 

Although the cubic type bound seem to be the limit of our methods for Theorem \ref{th:1}, an improvement on our results have been obtained by Y. Filmus and H. Hatami \cite{Filmus} via rather different and interesting methods. However, an optimal bound of $O(d)$ still remains open and (if true) it would be quite interesting to prove. A counter-example, for linear type bound may also be very interesting depending on the type of construction.

\section{Preliminaries}\label{prelim} 
In this work, we use concepts from analysis over the discrete spaces, specifically, the hypercube $\{-1,1\}^n$. For a good introduction to this area and its application to complexity theory, we refer to the surveys of de Wolf and O'Donnell \cite{ryan,ronald}.  We also refer to \cite{buhrman_wolf} for a good introduction to the complexity measures of functions such as randomized, quantum and deterministic query complexity and their relation to more analytic concepts such as degree and approximate degree, etc. 
We denote by $[n]$ the set of integers between $1$ to $n$. 
It is well-known that any function $f:\{-1,1\}^n \rightarrow \R $ can be represented as a polynomial with real coefficients over the  Fourier-Walsh characters: 
\[ f(x)= \sum_{S\subseteq[n]} \hat{f}(S) \, \chi_S(x) ,\]
where $\chi_S(x)=\prod_{i\in S} x_i$. The degree of $f$ is defined as
\[ \deg(f):= \max_{S\subseteq[n]: \,  \hat{f}(S)\neq 0} |S| .\]

Also, one of the tools used in the proof is the well-known noise operator:
\begin{definition}
The noise operator with rate $\rho\in \R$ is given by
\[T_{\rho} f(x):=\sum_{S \subseteq [n]} \widehat{f}(S) \, \rho^{|S|}\chi_S(x). \]
\end{definition}
For $\rho\in [-1,1]$, there is an alternative characterization of $T_\rho f$ which is useful for us: consider a bivariate distribution over $(x,y)\in \{-1,1\}^n\times \{-1,1\}^n$ defined by choosing $x\in \{-1,1\}^n$ uniformly at random, and for each $i\in [n]$ (independently) setting $y_i=x_i$ with probability $(1+\rho)/2$, and $y_i=-x_i$ with the remaining probability. It is not too hard to see \cite{ryan,ronald} that the above distribution, denoted by $x\sim_\rho y$, is symmetric in $x$ and $y$ and that the operator $T_\rho$ satisfies
\[ T_{\rho}\, f(x)=\Ex_{y \sim_{\rho}x}[f(y)], \]
for $\rho\in [-1,1]$. This characterization has the following useful consequence: for $\rho\in [-1,1]$ we have $\|T_\rho(f)\|_\infty \leq \|f\|_\infty$ and in fact $\|T_\rho(f)\|_q \leq \|f\|_q$ for all $q\geq 1$.

\begin{notation}[Dirac delta]  A Dirac delta or a point mass at a point $t\in \R$ is a probability measure $\mu=\delta(x-t)$ satisfying $\mu(K)=1$ if $t\in K\subseteq \R$, and otherwise $\mu(K)=0$. A weighted sum of Dirac delta measures over $\R$ is called a discrete measure. 
\end{notation}

\section{Proof overview}\label{sec:proof-overview}
 The proof of Theorem \ref{th:1} is best understood by focusing on the special case of homogeneous polynomials. Recall that a function $f$ is called 
 homogeneous if all of $f$'s non-zero Fourier coefficients $\hat{f}(R)\neq0$ satisfy $|R|\leq \deg(f)=d$. In fact, for homogeneous functions
we can prove a better estimate:
\begin{theorem}\label{thm:homg} Let $f$ be a function $f:\{-1,1\}^n\rightarrow \R$ that is homogeneous of degree $d$. Then
\[ \Inf(f)=O( d^2  \, \|f\|_\infty) .\]
\end{theorem}
Theorem \ref{th:1} is proved by a slight tweaking of parameters in the proof of Theorem \ref{thm:homg} (which costs us a factor of $d$ in the bound), and using some properties of the Chebyshev polynomials. 
Since the essence of the argument is already present in the proof of Theorem \ref{thm:homg}, from now on we assume the function $f$ is homogeneous of degree $d$. 

To prove Theorem \ref{thm:homg}, we introduce an operator $\MS f$, defined for each $S\subseteq [n]$. The action of $\MS$ on a function is to keep the Fourier coefficients of the characters that have intersection size $1$ with $S$ intact, and to zero out the rest of the Fourier expansion. More precisely, the operator is defined as follows:
\begin{definition}
Let $S\subseteq [n]$. $\MS$ is a linear operator on the space of functions over the discrete cube $\{-1,1\}^n$ given  by
\[ \MS f (x) := \sum_{R: \, | R\cap S|=1} \hat{f}(R) \chi_R(x) .\]
\end{definition}
One nice feature of $\MS f$ is that $\Inf_i(\MS f)$ for $i\in S$ has a particularly useful form as shown below in Fact \ref{fact:MS}. Another important property of $\MS$ is the following:
\begin{proposition}\label{prop:main} For all $f:\{-1,1\}^n\rightarrow \R$ with $d=\deg(f)$ and for all $S$, we have 
\[ \| \MS f \|_\infty = O(  d \:\|f\|_\infty ). \]
\end{proposition}
The quantity $\| \MS\|_{\infty\rightarrow\infty}=\sup_{f\neq0} \frac{ \|\MS f\|_\infty}{\|f\|_\infty}$ in general could be quite large; however, the above proposition guarantees that this quantity is reasonably small if we restrict the supremum to the bounded degree functions. The main idea for proving this proposition is to view the action of $\MS$ as a convolution:
\begin{equation} \MS f(x)= f * P_S(x)= \Ex_{y\in \{-1,1\}^n } f(y) P_S(xy).\footnote{Here $xy\in \{-1,1\}^n$ is the coordinate wise product of $x$ and $y$, i.e. $(xy)_i=x_iy_i$.} \label{eqn:conv} \end{equation}
If we wanted equation (\ref{eqn:conv}) to hold for all functions $f$, the function $P_S$ would be uniquely determined from the definition of $\MS$. However, we shall 
use the freedom given by the fact that $\deg(f)\leq d$ to choose a better $P_S$. 
\begin{proposition}\label{prop:main2}
There exists a function $P_S:\{-1,1\}^n\rightarrow\R$ satisfying
\begin{enumerate}
\item[(i)] $\widehat{P_S}(\{i\})=1$ for $i\in S$,
\item[(ii)] $\widehat{P_S}(A)=0$ for all $A\subseteq S$ with $|A|=0$ or $2 \leq |A|\leq d$,
\item[(iii)] $P_S(x)=0$ whenever there exists $i\in S^c$ with $x_i\neq 1$,
\end{enumerate}
such that $\| P_S \|_1= O(d)$. 
\end{proposition}
Consider the Fourier expansion of the function $P_S$ guranteed by the above proposition. 
\begin{align*} 
\widehat{P_S}(R) &= \Ex_{x\in\{-1,1\}^n} \left[ P_S(x) \chi_R(x)\right] \\
&= \frac{1}{2^n} \sum_{x:\: x_i=1 \; \forall i\in S^c} \left[ P_S(x) \chi_{R}(x) \right] \\
&= \widehat{P_S}(R\cap S) , 
\end{align*}
where we use the fact that the sum is over $x\in \{-1,1\}^n$ with $x_i=1$ for all $i\in S^c$ to deduce that $\chi_R(x)=\chi_{R\cap S} (x)$. It follows from Proposition \ref{prop:main2} that $\widehat{P_S(R)}= 1_{|R\cap S|=1}$. Here $1_{R\cap S=\{i\}}$ is a function which is $1$ when $R\cap S=\{i\}$ and is otherwise zero.  Notice that now Proposition \ref{prop:main} follows quickly because
\[ \widehat{f * P_S} (R)= \widehat{f}(R)\, \widehat{P_S}(R)= \widehat{f}(R) 1_{|R\cap S|=1}= \widehat{\MS f} (R). \]
On the other hand, 
\[ \|f * P_S\|_\infty\leq \| P_S\|_{1} \, \|f\|_\infty = O(d  \, \|f\|_\infty) . \]
Having defined $\MS f$ and investigated its properties, the next step is to define a quantity that allows us to get a better handle on the total $L_1$ influence. This quantity is denoted by $\mathcal I_p(f)$, parametrized by $p\in[0,1]$ . Here, $p$ should be thought of as a probability parameter which would be inverse polynomially related to the degree of $f$ in our setting.
\begin{definition} For a set $A$, we let $S\leftarrow_p A$ be a random subset of $A$ formed by including each $e\in A$ to be in $S$ independently with probability $p$. More formally, for any set $U\subseteq A$
\[ \Pr_{S\leftarrow_p A}[ S=U] = p^{|U|} (1-p)^{|A|-|U|} . \]
\end{definition}
\begin{definition}  Let $f:\{-1,1\}^n\rightarrow \R$. We define
\begin{equation} \mathcal I_p(f) :=  \Ex_{S \leftarrow_p [n]} \left[ \sum_{i\in S} \Inf_i \big( \MS f \big) \right] .  \label{Ip:def} \end{equation}
\end{definition}
The main hope here is that $\mathcal I_p(f)$ would act as a proxy for $\Inf(f)$, while being more tractable quantity to work with. More precisely, we want the following 
sandwiching relationship to hold for some choice of $p$:
\begin{equation} \frac{\Inf(f)}{d^{O(1)}} \leq \mathcal I_p(f) \leq d^{O(1)} \|f\|_\infty . \label{eqn:sandwich}\end{equation} 
Notice that equation (\ref{eqn:sandwich}) would prove (some form of) Theorem \ref{thm:homg}. Thus, for the rest of this section we shall  exclusively focus on the proof of these inequalities.

 There are two inequalities in equation (\ref{eqn:sandwich}). The right hand side of the inequality, i.e. $\mathcal I_p(f) \leq d^{O(1)} \|f\|_\infty$, holds for any $p\in [0,1]$. This is because of the next proposition (which is in fact the main reason we defined $\mathcal I_p(f)$ originally).
\begin{proposition}\label{prop:hard}
For any $x\in \{-1,1\}^n$, there exists some $y\in \{-1,1\}^n$ such that
\[ \sum_{i\in S} \left | \partial_i \MS f(x)\right| \leq  \MS f(y) . \]
\end{proposition}
We find the above proposition in some respects rather remarkable as it relates a (large) sum over the derivatives of a function to the value of the function itself (evaluated possibly at some other point of the discrete cube). Let us see how this proposition implies the right hand side of equation (\ref{eqn:sandwich}): 
\begin{align*} \sum_{i\in S} \Inf_i(\MS f) = \Ex_{x\in \{-1,1\}^n} &\left[ \sum_{i\in S}  \Big |\partial_i \MS f(x) \Big | \right]  \\ 
&\leq \| \MS f \|_\infty .\end{align*}
Hence, by the definition of $\mathcal I_p(f)$ and Proposition \ref{prop:main}, it follows that
\begin{align*} \mathcal I_p (f) &= \Ex_{S \leftarrow_p [n]}\left[ \sum_{i\in S} \Inf_i(\MS f)\right] \\
&\leq \max_S\| \MS f \|_\infty \\
&= O(d\, \|f\|_\infty) .\end{align*}
Hence, we proved the right hand side of equation (\ref{eqn:sandwich}). 
 
Let us now move on to the left hand side of equation (\ref{eqn:sandwich}). The main intuition here is that for a typical pair of $S\leftarrow_p [n]$ and $i\in S$, we would have 
\[ \Inf_i(f) \approx \Inf_i (\MS f). \]
Assuming this and recalling that for typical $S$ we have $|S|\approx pn$ (which should be thought of as the same order as $n$), it would be reasonable to expect that $\Inf(f)$ and $\sum_{i\in S} \Inf_{i}(\MS f)$ are closely related. This intuition is in fact correct in the sense that we have:
\begin{lemma} \label{lem:lower} Suppose $f:\{-1,1\}^n \rightarrow \R$ is a homogeneous of degree $d$. Then
\[ \mathcal I_p(f)\geq p(1-p)^{d-1} \Inf(f) .\]
\end{lemma}
We shall instantiate this lemma with $p=\frac{1}{d}$, which is chosen to (roughly) minimize $p(1-p)^{d-1}$. To prove this lemma we need the following fact:
\begin{fact}\label{fact:MS}
Let $S \subseteq [n]$ and $i\in S$. 
\[ \Inf_i \left( \MS f \right)= \Ex_{x} \left | \sum_{R\subseteq [n]} 1_{R\cap S= \{i\}} \: \hat{f}(R) \: \chi_{R}(x) \right | .\]
\end{fact}
The proof of this fact follows from the definition of $\MS$ and is straightforward. For completeness, a proof is given at the end of the section.

\begin{proof}[Proof of Lemma \ref{lem:lower}]
The plan is to swap the expectation $\Ex_{S\leftarrow_p [n]}$ and $\sum_{i\in S}$ in 
equation (\ref{Ip:def}). We can do this by fixing $i\in [n]$ and condition on the event $i\in S$ which occurs with probability $p$. Conditioned 
on this event, we have $S=S' \cup \{i\}$ with $S'\leftarrow_p [n]\setminus \{i\}$. Hence,
\[ \mathcal I_p(f) = p \sum_{i=1}^n \Ex_{S'\leftarrow_p [n]\setminus \{i\} } \left[\Inf_i\left(\Ms_{S'\cup \{i\}} f\right)  \right] , \]
where the term $p$ came from conditioning on the event $i\in S$. Using Fact \ref{fact:MS} in the above gives us
\[ \mathcal I_p(f) = p \sum_{i=1}^n \Ex_{\substack{S'\leftarrow_p [n]\setminus \{i\} \\x\in \{-1,1\}^n }}  
\left[   \bigg | \sum_{R \ni i} 1_{ R\cap S'= \emptyset} \: \hat{f}(R) \: \chi_{R }(x) \bigg |  \right] .\]
Note that since $S=S\cup \{i\}$, we translated $S\cap R=\{i\}$ to $i\in R$ and $S'\cap R=\emptyset$. Noting that $| R\setminus \{i\}| =d-1$ for all $R$ with $\hat{f}(R)\neq 0$, we have for such $R$'s
\begin{equation} \Ex_{S'\leftarrow_p [n]\setminus \{i \} }\left[ 1_{R\cap S'=\emptyset} \right] = (1-p)^{d-1} . \label{eqn:jel} \end{equation}
Now we use the triangle inequality to swap $| \cdot |$ and $\Ex_{S'\leftarrow_p [n]\setminus \{i\}}$. Substituting $(1-p)^{d-1}$  using equation (\ref{eqn:jel}), we get
\[ \mathcal I_p(f)  \geq  p \Ex_{x} \sum_{i=1}^n \left[   \bigg | \sum_{R:\: i\in R} (1-p)^{d-1} \: \hat{f}(R) \: \chi_{R}(x) \bigg |\right] ,  \]
which is precisely what we wanted to show.
\end{proof}
Hence, we have proved both sides of our central equation (\ref{eqn:sandwich}), finishing the proof of Theorem \ref{thm:homg}, except for the proof of Propositions \ref{prop:main} and \ref{prop:hard} given in Section \ref{sec:homg}. 

To prove Theorem \ref{th:1}, the main thing that must be modified is the statement of Lemma \ref{lem:lower}. There, the proof crucially depended on the fact that 
\[ \Ex_{S'\leftarrow_p [n]\setminus \{i \} }\left[ 1_{R\cap S'=\emptyset} \right] = (1-p)^{d-1}   \]
independently of $R$, which allowed us to take this term out of the expectation. When $f$ is not homogeneous, the above term, which is
\[\Ex_{S'\leftarrow_p [n]\setminus \{i \} }\left[ 1_{R\cap S'=\emptyset} \right] = (1-p)^{|R|-1},\]
cannot be pulled out of the expectation. The main trick is to apply the noise operator to $f$ before going through the computation of Lemma \ref{lem:lower}. More precisely, instead of working with equation (\ref{eqn:sandwich}), we work with a slightly different inequality:
\begin{equation}
\label{eqn:sandwich2} 
p \Inf(f) \leq \mathcal I_p ( T_{(1-p)^{-1}}  f ) \leq d^{O(1)} \|f\|_\infty . 
\end{equation}
Going through the same computation as that of Lemma \ref{lem:lower} with $T_{(1-p)^{-1}} f$ instead of $f$, allows us to prove the left hand side of equation (\ref{eqn:sandwich2}) with no modification. For the right hand side of equation (\ref{eqn:sandwich2}), we just need some facts about Chebyshev polynomials, specifically some estimates for $\| T_{(1-p)^{-1}} f \|_\infty$ in terms of $\|f\|_\infty$ and $p$. Notice that we are applying the noise operator with a rate $(1-p)^{-1}$, which is larger than one, and so $\| T_{(1-p)^{-1}} f \|_\infty$ could be much larger than $\| f\|_\infty$. Thus $p$ must be chosen well for this estimate to be useful.
\begin{proof}[Proof of Fact \ref{fact:MS}]
By definition
\[
\Inf_{i}\left( \MS f\right) = \Ex_{x\in \{-1,1\}^n} \left | \sum_{R \ni i: \,|R\cap S|=1} \hat{f}(R)\: \chi_R(x) \right|  . \label{eqn:fact}
\]
However, if $|R\cap S|=1$ and $i\in R, S$, the above sum is over sets $R$ with $|R\cap S|=\{i\}$. Hence,
\[ \Inf_i \left( \MS f \right)=\Ex_{x} \left | \sum_{R: \, R\cap S=\{i\} } \hat{f}(R)\:  \chi_{R}(x) \right| \] holds.
In the above expression, all $\chi_{R}(x)$ have $x_i=\pm 1$ as a common factor. Hence, we have the freedom to replace $R$ with $R\setminus \{i\}$ in the above expression, as we do elsewhere.
\end{proof}

\section{The case of homogeneous polynomials}\label{sec:homg}
As mentioned in Section \ref{sec:proof-overview}, the plan is to prove Theorem \ref{thm:homg} by proving the following two inequalities:
\[ \Inf(f)= O\big( d \: \mathcal I_{\frac{1}{d}} ( f) \big)= O\big( d^2  \, \|f\|_\infty \big) . \]
Setting $p=\frac{1}{d}$ in Lemma \ref{lem:lower} gives one of the two inequalities; the second inequality follows from a combination of Propositions \ref{prop:hard}, which gives
\begin{align*} \mathcal I_p (f)&= \Ex_{S \leftarrow_p [n]}\left[ \sum_{i\in S} \Inf_i(\MS f)\right]\\
&=\Ex_{x\in \{-1,1\}^n} \left[ \sum_{i\in S} \Big |\partial_i \MS f(x) \Big | \right]   \\ &
\leq \| \MS f \|_\infty ,\end{align*}

and Proposition \ref{prop:main}, which gives $\| \MS f \|_\infty=O(d\, \|f\|_\infty)$.

Let us first prove Proposition \ref{prop:hard}. 
\begin{proof}[Proof of Proposition \ref{prop:hard}]
Fix $x\in \{-1,1\}^n$ and $i\in S$. By the definition of $\MS f$, we have
\begin{align} \partial_i \MS f(x) &= \sum_{R: R\cap S=\{i\}} \hat{f}(R) \chi_R(x)  \\
&= x_i  \sum_{R: R\cap S=\{i\}} \hat{f}(R) \chi_{R\setminus\{i\}}(x).  \label{eqn:prop} \end{align}
Notice that since $x_i=\pm1$, and we are interested in the sum of the absolute value of the above expression, i.e. $\sum_{i\in S} \left |\partial_i \MS f(x)\right|$, the term $x_i$ can be dropped from the left hand side of equation (\ref{eqn:prop}). Define $y\in \{-1,1\}^n$ by
\[ y_i= 
\begin{cases} 
x_i  & \mbox{if}  \quad  i\notin S. \\
\sgn\left( \sum_{R:\, R\cap S=\{i\}} \hat{f}(R) \, \chi_{R\setminus\{i\}}(x)\right)   & \mbox{if} \quad i\in S. 
\end{cases}
\]
Notice that this choice of $y$ implies
\begin{align*} \sum_{i\in S} \left |\partial_i \MS f(x)\right| &=   \sum_{i\in S} \sum_{R: \; R\cap S=\{i\}} \hat{f}(R) \chi_{R}(y)\\
&= \sum_{R: \, |R\cap S|=1} \hat{f}(R)  \chi_{R}(y);\end{align*}
but the last term is precisely $\MS f(y)$. 
\end{proof}

We need some new definitions in order to prove Proposition \ref{prop:main}.
\begin{definition}
A measure $H$ supported on $B=[-1,1]$ is called $d$-admissible if  it satisfies the following conditions:
\begin{enumerate}\label{def:admis}
\item[(i)] $\int_B \gamma^{m}\; dH(\gamma)=0$ for $m=0$ and $2\leq m\leq d$, 
\item[(ii)] $\int_B \gamma\; dH(\gamma)=1$. 
\end{enumerate}
\end{definition}
\begin{remark}
\item[1.]The measures in our case are discrete, i.e. they consist of a weighted sum of point masses  as $H= \sum_{i=1}^m w_i \, \delta(\gamma- \alpha_i)$. This means for any $A\subseteq \R$, $H(A)=\sum_{\alpha_i \in A} w_i$.  For a discrete measure $H$, we define its absolute value, which is itself a measure with the same support as $H$, by
\[|H|:= \sum_{i=1}^m |w_i| \delta(\gamma-\alpha_i).\] 
\item[2.]We also define $\| H\|_1:= |H|(\R)= \sum_{i=1}^m |w_i|$. 
\end{remark}
The main lemma regarding the $d$-admissible measures we need is the following result proved in the appendix. 
\begin{lemma}\label{lem:vander}
For any $d\geq 1$, there exists a $d$-admissible measure as in Definition \ref{def:admis} with $\| H\|_1= O(d)$.\footnote{As noted in improved bounds and subsequent work part of Section \ref{sec:related_work}, the earlier versions of this paper only showed the existence a $d$-admissible measure with $\|H\|_1=O(d\log d)$ which caused the final bound to suffer by a logarithmic factor accordingly.}
\end{lemma}
Lemma \ref{lem:vander} in turn can be used to prove the next lemma which finishes the proof of Proposition \ref{prop:main}, and hence the proof of Theorem \ref{thm:homg}.

\begin{lemma}\label{lem:easy} For any $d$-admissible measure $H$ on $[-1,1]$, we can construct a function $P_S:\{-1,1\}^n\rightarrow \R$ satisfying the conditions of Proposition \ref{prop:main} with $\| P_S\|_1 \leq \| H\|_1= \int d |H(\gamma)|$.
\end{lemma}
\begin{proof}
Consider $P_S:\{-1,1\}^n\rightarrow \R$, specified as
\begin{equation} P_S(x)= \prod_{i \notin S} (1+x_i) \int_{-1}^1\prod_{i\in S} (1+\gamma x_i)\, d H(\gamma) ,  \label{eqn:lem} \end{equation}
which can be easily seen to satisfy property (iii) in Proposition \ref{prop:main}. Also, notice that $\widehat{P_S}(\{i\})$ for $i\in S$ is exactly the coefficient of the monomial $x_i$ in equation (\ref{eqn:lem}). By our guarantee on $H$'s first moment, we have $\widehat{P_S}(\{i\})=\int_{-1}^{1} \gamma\: d H(\gamma)= 1$ for $i\in S$. Similarly for computing the Fourier coefficient for $\widehat{P_S}(A)$ for some $A\subseteq S$, we need to see what is the term we pick up from the second product in equation (\ref{eqn:lem}). In general, we see that  for any $A\subseteq S$
\[ \widehat{P_S}(A)= \int_{-1}^{1} \gamma^{|A|} \, dH(\gamma) .  \]
A moment of reflection reveals that actually for any $A\subseteq [n]$,
\[ \widehat{P_S}(A)= \int_{-1}^{1} \gamma^{|A\cap S|} \, dH(\gamma) \,   \]
holds. This, combined with the properties guaranteed on moments of $H$, proves that the proposed $P_S$ in equation (\ref{eqn:lem}) satisfies the conditions of Proposition \ref{prop:main}. We are just left with the computation of $\|P_S\|_1$ which is 
\[
 \Ex_{x\in \{-1,1\}^n} \left[ \prod_{i \notin S} |1+x_i| \:  \Big |\int_{-1}^1\prod_{i\in S} (1+\gamma x_i)\, d H(\gamma) \Big | \: \right].
\]
For simplicity assume $S=[k] \subseteq [n]$. Notice that $|1+x_i|= 1+x_i$, and hence $\Ex[|1+x_i|]=1$. Using independence of the random variables $\{x_i\}_{i\in S}$, and the triangle inequality, we see that 
\begin{align*}
\| P_S\|_1 &= \Ex_{x\in \{-1,1\}^k} \left[   \Big |\int_{-1}^1\prod_{i=1}^{k} (1+\gamma x_i)\, d H(\gamma) \Big | \right] \, \\
  & \leq   \Ex_{x\in \{-1,1\}^k} \left[   \int_{-1}^1\prod_{i=1}^{k} |1+\gamma x_i|\, d| H(\gamma)|  \right] \, \\
  & =  \int_{-1}^1 d |H(\gamma)| \, \prod_{i=1}^{k}\,  \Ex_{x_i} |1+\gamma x_i | =\int_{-1}^{1} d |H(\gamma)|,
\end{align*}
which is $\|H\|_1$. Here, we used the fact that $|1+\gamma x_i|= 1+ \gamma x_i$ for $\gamma\in [-1,1]$ and $\Ex[x_i]=0$. The above is exactly our desired result. 
\end{proof}

\section{The general case}\label{sec:general}
The steps are completely analogous to the homogeneous case which we discussed in detail in Section \ref{sec:proof-overview} and \ref{sec:homg}. See the above sections for more explanations of the arguments. 
\begin{lemma} Let $f:\{-1,1\}^n\rightarrow \R$ of degree $d$. We have
\[ p \Inf(f) \leq \mathcal I_p( T_{(1-p)^{-1}} f) .\]
\end{lemma}

\begin{proof}
Using the definitions of $\mathcal I_p(\cdot)$ and the noise operator, and a  triangle inequality we get
\begin{align*}
& \mathcal I_p(T_{(1-p)^{-1}}  f) \\ 
&= p \sum_{i=1}^n \Ex_{S'\leftarrow_p [n]\setminus \{i\} } \left[\Inf_i\left(\Ms_{S'\cup \{i\} } T_{(1-p)^{-1}} f\right)  \right] \\
  &= p \sum_{i=1}^n \Ex_{x, S'}\left| \bigg[ \sum_{R \ni i} (1-p)^{-|R|} 1_{R\cap S'=\emptyset}  \hat{f}(R) \chi_R \bigg] \right|  \\ 
  &\geq p \sum_{i=1}^n  \Ex_{x}\left| \bigg[ \sum_{R \ni i}\hat{f}(R) \chi_R (1-p)^{-|R|} \Ex_{S'\leftarrow_p [n]\setminus \{i\}}[1_{S'\cap R}] \bigg] \right| \\
  &= \frac{p}{1-p} \sum_{i=1}^n  \Ex_{x}\left|\bigg[   \sum_{R \ni i}\hat{f}(R) \chi_R \bigg] \right| \geq p \, \Inf(f) .
 \end{align*}

\end{proof}
As a consequence of Proposition \ref{prop:main} we have 
\begin{equation} \mathcal I_p (T_{(1-p)^{-1}} f )= O(d  \| T_{(1-p)^{-1}} f\|_\infty) . \label{eqn:general}
\end{equation}
So we need to estimate $\| T_{(1-p)^{-1}} f \|_\infty$ in term of $\|f\|_\infty$. Consider the point $x\in\{-1,1\}^n$ such that $| T_{(1-p)^{-1}} f(x)| = \|  T_{(1-p)^{-1}} f \|_\infty$. Fix $x$ and view $T_{\rho} f(x)$ as a univariate polynomial of degree $d$ in $\rho$. Recall from the preliminaries that $|T_\rho f(x) | \leq \|f\|_\infty$ for all $\rho\in [-1,1]$. The crucial lemma is the following which is a well-known fact from approximation theory. 
\begin{lemma}
Suppose $P$ is a polynomial of degree $d$. Then for $x>1$,
\[ |P(x)|\leq e^{2d\sqrt{x^2-1}} \max_{t\in [-1,1]} |P(t)| .\]
\end{lemma}
\begin{proof}
By scaling we can assume $|P|\leq 1$ for $t\in [-1,1]$. Classical properties of Chebyshev polynomials \cite{rivlin} indicate that $|P(x)| \leq T_d(x)$ where $T_d$ is d\textsuperscript{th} Chebyshev polynomial. We have
\[ T_d(x)= \frac{1}{2} \left((x+\sqrt{x^2-1})^{d} + (x- \sqrt{x^2-1})^d \right). \]
For $x\geq 1$, $\sqrt{x^2-1}\geq x-1$ and hence
\[ T_d(x) \leq (x+ \sqrt{x^2-1})^d \leq (1+ 2\sqrt{x^2-1})^d \leq e^{2d\sqrt{x^2-1}}. \]
\end{proof}
Setting $p=\frac{c}{d^2}$ for some $c>0$, and estimating $(1-p)^{-2} -1 = \Theta(c / d^2)$ gives us
\[ \Inf(f)= O(c^{-1} e^{\Theta(\sqrt{c})} d^3  \|f\|_\infty ) .\]
This finishes the proof of Theorem \ref{th:1}. 

\section{A corollary on maximal deviation of cut-value of graphs }\label{sec:app}
In this section we use a special case of Theorem \ref{th:1} to give a new proof of a theorem of Erd\H{o}s et al. \cite{erdos} on the maximum discrepancy of cut-values in (unweighted) graphs. Given $S\subseteq V$ a vertex subset of a graph $G=(V,E), $ we denote $V\setminus S$ by $S^c$ and we write $u \sim v$ if and only if $(u,v) \in E$.

\begin{definition}
For any graph $G=(V,E)$ and $0 \leq p \leq 1$ the cut-deviation $D_p(G)$ is the maximum over all cuts $(S, V\backslash S)$ of the discrepancy between the cut-value $\left| E(S,S^c)\right|$ and the expected cut-value $p |S| (|V|-|S|)$ (where we choose each edge independently with probability $p$), i.e.,
\[  D_p(G) = \max_{S\subseteq V} \Big | \big|E(S,S^c)\big|- p |S||S^c| \Big |. \]
\end{definition}
We are interested in lower bounding the quantity $D_p(G)$. For a $G=(V,E)$, let $\rho_G:=|E|/\binom{|V|}{2}$ denote the edge density of $G$. Notice that for any $p\neq \rho_G$ a random cut will already give a deviation of $\Omega(n^2)$ for $D_p(G)$. So the interesting case is when $p=\rho_G$.  For this choice of the parameter we prove the following theorem:
\begin{theorem} \label{thm:erdos} For every graph $G=(V,E)$,  \label{graph:cor}
\[  D_{\rho_G}(G)= \Omega(\min(\rho_G,1-\rho_G)n^\frac{3}{2}). \]
\end{theorem}
We note that the above inequality is tight which can be seen by applying Chernoff bound to Erd\H{o}s-Renyi graphs $\mathcal G(n,p)$. More formally, if $G\sim \mathcal G(n,p)$ for all $S\subset V$ and some $c,C>0$, which may depend on $p$, we have 
\begin{align*} \Pr\Big[ &\big|E(S,S^c)-p |S||S^c| \big| \geq \alpha n\Big]  \\
 & \leq  \Pr\Big[ \big |E(S,S^c)-p |S||S^c | \big|  
\geq \alpha \sqrt{|S| |S^C|}\Big] \\
&\leq c \exp(-C \alpha^2) . \end{align*}
Taking $\alpha= r \sqrt{n}$ for appropriate constant $r$ in the above, and applying a union bound over all cuts $S\subset V$, proves the tightness of Theorem \ref{thm:erdos}. Moreover, the one-sided variant of this inequality
\[ \max_{S\subseteq V} E(S,S^c)- \rho_G |S| |S^c| = \Omega(\min(\rho_G, 1-\rho_G) n^\frac{3}{2}) \]
 which holds for random graphs, does not hold in general; this can be seen from the example of the complement of complete bipartite graph $K_{n/2, n/2}$. Thus, the result is optimal in this sense as well.

To prove Theorem \ref{thm:erdos},  we use the following:
\begin{lemma}
Let $G=(V,E)$ with $|V|=n$, and assume $V=[n]$. We associate to any $S \subseteq V$ a point $x\in \{-1,1\}^n$ by setting $x_i=1$ for $i \in S$ and $x_i=-1$ for $i\in S^c$. 
Then
\[ g_p(x):=\frac{|E|}{2}-p\frac{|V|(|V|-1)}{4}+\frac{p}{2} \sum_{i<j} \,x_i x_j - 1/2 \sum_{i\sim j}\,  x_i x_j \]
satisfies $g_p(x_S)=E(S, S^c) - p |S| |S^c|$.
\end{lemma}
\begin{proof}
This is a standard computation; one checks that $|E(S,S^c)|=1/2 |E|-1/2\sum_{i \sim j} x_i x_j$ and $|S||S^c|={|V|(|V|-1) \over 4}-{1 \over 2}\sum_{i<j} x_i x_j.$
\end{proof}
Now we are ready for our final proof:
\begin{proof}[Proof of Theorem \ref{thm:erdos}] Let $p=\rho_G$ since, as noted above, if $p\neq \rho_G$, a random cut achieves a $\Omega(n^2)$ lower bound. Notice that 
\[ \| g_p(G)\|_\infty = \max_{S\subseteq [n]} \left | E(S,S^c)- p|S||S^c|\right|=D_{p}(G), \]
where in the first equality we used the previous lemma. Theorem \ref{th:1} implies
\begin{align} \label{eqn:erdos} 
\Inf(g_{p}) &=\sum_{i=1}^n \, \Ex_x \bigg[  \Big| \frac{p}{2}\sum_{j\nsim i} x_j - \frac{1-p}{2}\sum_{j \sim i} x_j \Big| \bigg] \\
&= O\Big(  \max_{S\subseteq [n]} \big | E(S,S^c)- p|S||S^c|\big| \Big),  
\end{align}
where we use the fact that $\deg(g_{p})=2$.

Now we just need to a lower bound for the left hand side of the previous equation. Fix a particular $i^*\in V$. We claim  the expression
\[ \Ex_x \bigg[  \Big|  p/2\sum_{j\nsim i^*} x_j - (1-p)/2\sum_{j \sim i^*} x_j \Big| \bigg] \] 
is $\Omega\left(\min(p,1-p)\sqrt{n}\right)$ (in the above both sums are over $j$, for a fixed $i^*$). Note that at least one of the sums has at least $n/2$ terms. Without loss of generality assume, $|\{j\sim i^*\}| \geq n/2$. From the central limit theorem (or simple properties of binomial distributions) then follows that
\[  \Ex \bigg[ \Big| \sum_{j \sim {i^*}}  x_j \big| \bigg] =\Theta(\sqrt{n}) .\]
Now applying the Jensen's inequality to take $\Ex_{x_j}$ for $j\nsim i^*$ inside the expectation, and using $\Ex[ x_j]=0$, it follows that
\begin{align*} \Ex_x \bigg[  \Big|  \frac{p}{2}\sum_{j\nsim i^*} x_j - \frac{1-p}{2} \sum_{j \sim i^*}   x_j \Big|& \bigg] \geq \frac{1-p}{2} \Ex\bigg[ \Big| \sum_{j \sim i^*} x_j \Big| \bigg]\\
&=\Omega\left(\min(p,1-p)\sqrt{n}\right) .\end{align*}
\end{proof}

\section{Conclusion}\label{open_problem_section}
The main open problem is to improve the bound in Theorem \ref{th:1}. We believe that this bound is far from optimal.  It is  conceivable that  the total $L_1$ influence of a $[-1,1]$-valued function $p$ is always bounded by a linear function of the degree of $p$. 

As mentioned in the introduction, we hope that our results and techniques in this work would be useful in the study of quantum versus classical query complexity of partial functions. However,  as demonstrated in Section \ref{sec:app}, the applications of our inequality may not be limited to complexity theory. There, we gave a proof of a purely combinatorial result of Erd\H{o}s et al. by applying Theorem \ref{th:1} to an appropriately chosen polynomial.

Another possible future direction is to clarify the relationship between the notion of $L_1$ influence in the discrete cube as studied in this work and the alternative notions of $L_1$ type influences in the Gaussian setting as discussed in  \cite{ledoux, geometric_1, geometric_2} and also the recent one in \cite{feldman}.
\section*{Acknowledgments}

We thank Scott Aaronson for introducing this problem to us. We would also like to thank Jelena Markovic, Madhu Sudan and Yuval Filmus for helpful comments and suggestions which led to much improvements to the exposition. MB was partially supported by National Science Foundation through STC-award 0939370. 

\small
\singlespacing
\bibliographystyle{siam}

\appendix
\section{Construction of optimal $d$-admissable measure on $\R$}
In this section we prove Lemma \ref{lem:vander}. Without loss of generality we can assume that the degree parameter $d \geq 1$ is an odd integer since this can be guaranteed by increasing $d$ to $d+1$; the only negative effect of this is a slight worsening of the hidden constant in $O(\cdot)$ in the conclusion of Lemma \ref{lem:vander}.

We start by showing that the resulting measure is optimal (as such improving this lemma cannot be used to make much further progress in the bounds obtained in Theorem \ref{th:1}).
To this end, we consider the Chebyshev polynomial $T_d(\gamma)$.

\begin{theorem} \label{thm:lb}
 If $H$ is $d$-admissible then $\|H\|_1 \geq d$.
 Furthermore, equality is only possible for the measures supported on
 $\gamma_t = \cos \tfrac{t\pi}{d}$ for $0 \leq t \leq d$.
\end{theorem}
\begin{proof}
 Since $|T_d(\gamma)| \leq 1$, we have
\[
 \int_{-1}^1 \intd|H(\gamma)| \geq \left|\int_{-1}^1 T_d(\gamma) \, \intd
 H(\gamma)\right| = d,
\]
 since $T_d$ has degree $d$ and $\left.T_d(\gamma)\right|_{\gamma^0} = 0$ while $\left.T_d(\gamma)\right|_{\gamma^1} = (-1)^{(d-1)/2} d$.

 Equality is only possible for a measure concentrated on the values
 satisfying $|T_d(\gamma)| = 1$. If $\gamma = \cos \theta$ then
 $T_d(\gamma) = \cos (d\theta)$, and so $d \theta = t \pi$ for some
 integer $t$.
\end{proof}

The measure we construct will actually be supported only on $\gamma_1,\ldots,\gamma_d$.
For the rest of this section, we consider the atomic measure supported
on $\gamma_1,\ldots,\gamma_d$ and given by
\begin{equation} \label{eq:def}
H(\{\gamma_t\}) = \frac{(-1)^{(d-1)/2+t}}{d} \cdot \begin{cases} 1, & t = d, \\ \gamma^{-2}
  - \gamma^{-1}, & t < d. \end{cases}
\end{equation}

We first show that $H$ is $d$-admissible. We start by giving a formula
which will help us calculate the required integrals.

\begin{lemma} \label{lem:formula}
We have
\[
 (-1)^{(d-1)/2} d \int_{-1}^1 \gamma^k \, \intd H(\gamma) = (-1)^{k+1} + S_{k-2} - S_{k-1},
\]
where
\[
 S_\ell = \sum_{t=1}^{d-1} (-1)^t \gamma_t^\ell.
\]
\end{lemma}
\begin{proof}
 Follows straight from the definition of $H$, using $\gamma_d = \cos \pi = -1$.
\end{proof}

We proceed to calculate $S_\ell$ for the relevant values of $\ell$,
namely $-1 \leq \ell \leq d-2$.

\begin{lemma} \label{lem:even}
 When $\ell$ is even, $S_\ell = 0$.
\end{lemma}
\begin{proof}
 Since $\cos (\pi - \theta) = -\cos \theta$, we have $\gamma_t^\ell =
 \gamma_{d-t}^\ell$ and so $(-1)^t \gamma_t^\ell + (-1)^{d-t}
 \gamma_{d-t}^\ell = 0$.
\end{proof}

\begin{lemma} \label{lem:odd}
 When $\ell$ is odd,
\[ S_\ell = -1 + \sum_{t=1}^d (\cos \tfrac{2 t \pi}{d})^\ell. \]
\end{lemma}
\begin{proof}
 Using $\cos(\pi - \theta) = -\cos \theta$, we have
\begin{align*}
 S_\ell &= \sum_{t=1}^{d-1} (-1)^t (\cos \tfrac{t\pi}{d})^\ell \\ &=
 \sum_{t=1}^{(d-1)/2} (\cos \tfrac{2t\pi}{d})^\ell +
 \sum_{t=1}^{(d-1)/2} (- \cos \tfrac{(2t-1)\pi}{d})^\ell \\ &=
 \sum_{t=1}^{(d-1)/2} (\cos \tfrac{2t\pi}{d})^\ell +
 \sum_{t=1}^{(d-1)/2} (\cos \tfrac{(d-2t+1)\pi}{d})^\ell \\ &=
 \sum_{t=1}^{d-1} (\cos \tfrac{2t\pi}{d})^\ell. \qedhere
 \end{align*}
\end{proof}

In order to compute $S_\ell$, we employ the residue calculus.

\begin{lemma} \label{lem:residue}
 When $\ell \leq d-2$ is odd,
 \[
 S_\ell = -1-\sum_{w\colon w^d \neq 1} \Res(f(z),z=w), \]
where $f(z)=\frac{d(z^2+1)^\ell}{2^\ell z^{\ell+1}(z^d-1)}$, where the sum is over all poles of $f(z)$ other than those at $d$th
 roots of unity.
\end{lemma}
\begin{proof}
 Consider the function
\[ f(z) = \left(\frac{z+z^{-1}}{2}\right)^\ell \frac{d}{z^{d+1}-z}. \]
 When $|z|$ is large, $|f(z)| = O(z^{\ell-d-1}) = O(z^{-3})$, and so
 if we integrate $f(z)$ over a large circle around the origin, the result will be
 $O(z^{-2})$ and so will tend to zero as the radius tends to
 infinity. On the other hand, the residue theorem implies that the
 integral equals the sum of residues of the function, over $2\pi
 i$. We conclude that the sum of residues equals zero. It is
 well-known that the function $d/(z^{d+1}-z)$ has residue $1$ at $d$th
 roots of unity, and so
\[ \Res(f(z),z=\exp \tfrac{2 t \pi i}{d}) = \left(\frac{z+z^{-1}}{2}\right)^\ell
= (\cos \tfrac{2t \pi}{d})^\ell. \]
 The lemma now follows from Lemma~\ref{lem:odd}.
\end{proof}

Using the formula obtained in the preceding lemma, we calculate
$S_\ell$ for odd $\ell$, separately for $\ell \geq 1$ and $\ell = -1$.

\begin{lemma} \label{lem:odd-easy}
 When $1 \leq \ell \leq d-2$ is odd, $S_\ell = -1$.
\end{lemma}
\begin{proof}
 Using Lemma~\ref{lem:residue}, we have to compute the residue of the
 following function at $z = 0$:
\[ f(z) = \frac{d(z^2+1)^\ell}{2^\ell z^{\ell+1}(z^d-1)}. \]
 Around $z = 0$ we have
\[
 f(z) = - \frac{d(z^2+1)^\ell}{2^\ell z^{\ell+1}} (1 + O(z^d)).
\]
 Opening the binomial coefficient, since all powers of $z$ are even,
 we see that the coefficient of $z^{-1}$ is zero.
\end{proof}

\begin{lemma} \label{lem:odd-hard}
 We have $S_{-1} = -1 + (-1)^{(d-1)/2} d$.
\end{lemma}
\begin{proof}
 Using Lemma~\ref{lem:residue}, we have to compute the residues of the
 following function at the ``non-trivial'' poles $z = \pm i$:
\[ f(z) = \frac{2d}{(z^2+1)(z^d-1)}. \]
 Since the poles are simple, it is easy to compute
\begin{align*}
 \Res(f(z),z=i) + \Res(f(z), z=-i &) = \frac{2d}{2i (i^d-1)} \\ + &
 \frac{2d}{(-2i) (-i^d-1)}.
\end{align*}
 When $(d-1)/2$ is even,
\begin{align*}
 \Res(f(z),z=i) + \Res(&f(z),z=-i ) = \frac{d}{i(i-1)}  \\
 & +
 \frac{d}{(-i)(-i-1)} = -d.
\end{align*}
Therefore, in this case $S_{-1} = -1 + d$. Similarly, when $(d-1)/2$
is odd the sum of residues is $d$, and so $S_{-1} = -1-d$.
\end{proof}

The preceding two lemmas, together with Lemma~\ref{lem:formula} and Lemma~\ref{lem:even}, allow
us to prove that $H$ is $d$-admissible.

\begin{lemma} \label{lem:admissible}
 The measure $H$ is $d$-admissible.
\end{lemma}
\begin{proof}
 We start with property~(i). Lemma~\ref{lem:formula} together with
 Lemma~\ref{lem:odd-hard} and Lemma~\ref{lem:even} show that
\[
 (-1)^{(d-1)/2}d \int_{-1}^1 \gamma \, \intd H(\gamma) = 1 + S_{-1} - S_0 = (-1)^{(d-1)/2}d.
\]

 Property~(ii) is similar. Suppose first that $2 \leq d \leq k$ is
 even. Lemma~\ref{lem:formula} together with
 Lemma~\ref{lem:odd-easy} and Lemma~\ref{lem:even} show that
\begin{align*}
 (-1)^{(d-1)/2}d \int_{-1}^1 \gamma^k \, \intd H(\gamma) &= -1 + S_{d-2} -
 S_{d-1} \\ &= -1 + 0 - (-1) = 0.
\end{align*}
 Similarly, when $2 \leq d \leq k$ is odd we have
\begin{align*}
 (-1)^{(d-1)/2}d \int_{-1}^1 \gamma^k \, \intd H(\gamma) &= 1 + S_{d-2} -
 S_{d-1} \\
 &= 1 + (-1) -0 = 0. \qedhere
\end{align*}
\end{proof}

It remains to calculate $\|H\|_1$. We do this in two
steps.

\begin{lemma} \label{lem:formula-2}
 We have
\[
 \|H\|_1 = \frac{1}{d} \sum_{t=1}^d \frac{1}{(\cos \tfrac{\pi k}{d})^2}.
\]
\end{lemma}
\begin{proof}
 By definition,
\begin{align*}
 d\|H\|_1 &= d \int_{-1}^1 \intd|H(\gamma)| = d \sum_{t=1}^d |H(\{\gamma_t\})| \\
 &= 1 +
 \sum_{t=1}^{d-1} |\gamma_t^{-2} - \gamma_t^{-1}|.
\end{align*}
 We claim that $\gamma_t^{-2} > \gamma_t^{-1}$ for all $1 \leq t \leq
 d-1$. If $\gamma_t$ is negative, this is clear. If $\gamma_t$ is
 positive then since $\gamma_t < 1$, clearly $\gamma_t^{-1} <
 \gamma_t^{-2}$. Therefore
\[
 d\|H\|_1 = 1 + \sum_{t=1}^{d-1} \gamma_t^{-2} -
 \sum_{t=1}^{d-1} \gamma_t^{-1}.
\]
 Since $\gamma_t = -\gamma_{d-t}$, the second sum vanishes, and we
 conclude
\[
 d \|H\|_1 = 1 + \sum_{t=1}^{d-1} \gamma_t^{-2} =
 \sum_{t=1}^d (\cos \tfrac{\pi k}{d})^{-2}. \qedhere
\]
\end{proof}

We can evaluate the sum using the residue calculus.

\begin{lemma} \label{lem:sum}
 We have
\[
 \sum_{t=1}^d \frac{1}{(\cos \tfrac{\pi k}{d})^2} = d^2.
\]
\end{lemma}
\begin{proof}
 Consider the function
\[ f(z) = \frac{4d}{(z+1)^2(z^d-1)} = \frac{4z}{(z+1)^2}
\frac{d}{z^{d+1}-z}. \]
 As in the proof of Lemma~\ref{lem:residue}, the sum of residues
 vanishes. The residue at a $d$th root of unity $z = \exp \tfrac{2t\pi
 i}{d}$ is
\begin{align*}
 \left. \frac{4z}{(z+1)^2} \right|_{z = \exp \tfrac{2t\pi
 i}{d}} = \left. \frac{2}{1 + (z + z^{-1})/2} \right|_{z = \exp \tfrac{2t\pi
 i}{d}} \\
 = \frac{2}{1 + \cos \tfrac {2t\pi}{d}} = \frac{1}{(\cos \tfrac{t\pi}{d})^2},
\end{align*}
 using the identity $2(\cos \theta)^2 = 1 + \cos (2\theta)$. In order
 to calculate the residue at $-1$, we calculate instead the residue of
 $g(w) = f(z+1)$ at $w = 0$:
\begin{align*}
 g(w) = \frac{4d}{w^2((w-1)^d-1)} &= \frac{4d}{w^2(-2+dw+O(w^2))} \\ &=
 \frac{-2d}{w^2(1-(d/2)w+O(w^2))} \\
 &= \frac{-2d}{w^2} (1 + \tfrac{d}{2}w
 + O(w^2)).
\end{align*}
 Therefore $\Res(f(z),z=-1) = \Res(g(w),w=0) = -d^2$. The formula
 immediately follows.
\end{proof}

Hence, the main result of the section which is the proof of Lemma \ref{lem:vander} follows.

\end{document}